\documentclass[a4paper, 12pt]{amsart}
\usepackage{graphicx}
\usepackage{amsfonts}
\usepackage{amsmath}
\usepackage{amssymb}

\begin{document}

\title{The magic of universal quantum computing with permutations}

\author{Michel Planat$\dag$}
\address{$\dag$Institut FEMTO-ST, CNRS and Universit\'e de Bourgogne/Franche-Comt\'e, 15 B Avenue des Montboucons, F-25044 Besan\c con, France.}
\email{michel.planat@femto-st.fr}

\author{Rukhsan-Ul-Haq$\ddag$}
\address{$\ddag$Jawaharlal Nehru Centre for Advanced Scientific Research, Jakkur Bengaluru, India. }
\email{mrhaq@jncasr.ac.in}

\begin{abstract}

The role of permutation gates for universal  quantum computing is investigated. The \lq magic' of computation is clarified in the permutation gates, their eigenstates, the Wootters discrete Wigner function and state-dependent contextuality (following many contributions on this subject). A first classification of  resulting magic states in low dimensions $d \le 9$ is performed.

\end{abstract}

\maketitle

\vspace*{-.5cm}
\footnotesize {~~~~~~~~~~~~~~~~~~~~~~PACS: 03.67.-a, 02.20.-a, 03.65.Fd, 03.65.Aa, 02.10.Ox, 03.65.Ud} 

\footnotesize {~~~~~~~~~~~~~~~~~~~~~~MSC codes: 81P68, 81P45, 20B05, 81P13}
\normalsize

\section{Introduction}

Quantum physics and (universal) computing are now considered to be deeply interrelated. This relatively new idea of science owes much to R. Feynman \cite{Feynman1982}, D. Deutsch \cite{Deutsch1985}, P. Shor \cite{Nielsen2000}, S. Bravyi and A. Kitaev \cite{Bravyi2004}, to mention a few popular landmarks. The time line of quantum computing also includes other important marks with many of them, during the last two decades, dedicated to quantum computing platforms \footnote{Let us mention Intel and NTT for spin qubits in semiconductors, Google, IBM and D-Wave for superconducting qubits, Lockheed and INFINEON for trapped ions. Other hardware efforts are being accomplished at university laboratories with linear optics, atoms and cavity QED, quantum dots and impurity spins in solids, etc \cite{QIP2005,Report2012}.}. R. Feynman already understood that the simulation of a quantum system on a classical computer would need exponential resources. Later D. Deutsch proposed a universal quantum computer made of quantum gates as a way of simulating a quantum system with at most a polynomial overhead. We know from P. Shor that a quantum computer is able to factor large integers 
in polynomial time by exploiting the parallelism in the implementation of quantum Fourier transform \cite[p. 221]{Nielsen2000} and that quantum error correction may circumvent the undesirable effects due to decoherence by the use of quantum error-correcting codes \cite[p. 435]{Nielsen2000}. S. Bravyi and A. Kitaev introduced the principle of \lq magic state distillation': universal quantum computation may be realized thanks to the stabilizer formalism (Clifford group unitaries, preparations and measurements \cite{Gottesman1998}) and the ability to prepare an ancilla in an appropriate single qubit mixed state. 

Within the frame of universal quantum computation based on the stabilizer formalism, it is being actively discussed if there exist critical resources responsible for the power of quantum computation. Remarkably, in odd dimensions, contextuality of the magic states seems to be the magic ingredient \cite{Howard2014}. In addition, the contextuality of states is witnessed by the negative entries of a quasi-probability distribution - the discrete Wigner function (DWT). For even dimensions, the situation is more obscure since state-independent contextuality (correlated with the negativity of DWT) occurs from stabilizer states. Filtering of the quantum states involved in the computation based on full state tomography has been proposed \cite{Raussendorf2016}. Finally, according to \cite{Spekkens2008}, contextuality is required not just for measurement procedures but for preparation procedures as well, in order that	the two notions of nonclassicality are revealed to be equivalent.

Wigner function was recognized to contain permutation symmetry in its structure \cite{Zhu2016}. Interestingly enough, the experimental implementation of a simple quantum algorithm for determining the parity of a permutation was performed \cite{Gedik2015}. We claim in this paper that permutation symmetry can sometimes be considered lying beneath the concepts of magic states and contextuality that are responsible for the universality and efficiency of quantum computation. Permutation gates in the Clifford group (whose important elements are the CNOT gate and the Toffoli or CCNOT gate) reveal non-stabilizer states in their eigenvectors.
 The focus of this contribution is the use of two-generator permutation groups seen as sets of permutation gates and being the source of quantum states (stabilizer or magic) arising from their eigenstates. The groups of interest are in general different from the ones encountered in the previous investigation of the Kochen-Specker theorem \cite{Planat2015}. Only state-dependent contextuality is involved in quantum computational universality, magic state distillation through quantum error-correcting codes and possibly computational speed-up, as already emphasized \cite{Howard2014}.

In Sec. 2, we remind standard results about the generalized Pauli group, the phase-point operators, Wootters discrete Wigner function and its link to quantum contextuality. In Sec. 3, we derive several types of magic states arising from permutation matrices in small dimensions $d \le 9$, we explicit their Wigner function and obtain state-dependent proofs of contextuality based on the existence of pentagons between the appropriate states (stabilizer and magic). Sec. 4 describes open vistas for further study.

\subsection*{From permutations to quantum gates}
It is unusual to recognize the relationship of permutations with quantum gates as we intend to do in this work. A $n$-letter permutation admits a $n \times n$ binary matrix representation with exactly one entry of $1$ in each row and each column and $0$s elsewhere. \lq Magic' permutation matrices are those showing one entry of $1$ on their main diagonals. Some well known permutation matrices/gates are the Pauli gate $X=\bigl( \begin{smallmatrix} 
  0 & 1\\
  1 & 0 
\end{smallmatrix} \bigr)\equiv (2,1)$, $I \otimes X\equiv (2,1)(4,3)$, CNOT=$\bigl(\begin{smallmatrix} 
  1 & 0&0&0\\
  0 & 1&0&0\\
	0&0&0&1 \\
	0&0&1&0
\end{smallmatrix} \bigr)\equiv (1,2)(4,3)$, CCNOT$\equiv(1,2,3,4,5,6)(8,7)$, that acts on one, two or three qubits, respectively. Similarly permutation gates may act on qudits as the shift gate $X=\bigl( \begin{smallmatrix} 
  0 & 1&0\\
  0&0 & 1 \\
	1&0&0
\end{smallmatrix} \bigr)\equiv (2,3,1)$ acting on qutrits.

In Sec. 3, we focus on magic groups generated by two magic permutation gates (they exist as soon as $d \ge 4$ as in (\ref{eqn11}) of Sec. 3.3).

\section{The generalized Pauli group, the discrete Wigner function and contextuality}
We remind standard tools useful for further calculations and discussion.
\label{Sec2}

\subsection*{The generalized Pauli group}
Let $d$ be a prime number, the qudit Pauli group is generated by the shift and clock operators as follows
\begin{eqnarray}
&X\left|j \right\rangle = \left|j+1 \mod d \right\rangle \nonumber \\
&Z \left |j \right\rangle=\omega^j \left|j \right\rangle 
\label{eqn1}
\end{eqnarray}
with $\omega=\exp(2i\pi/d)$ a $d$th root of unity. In dimension $d=2$, $X$ and $Z$ are the Pauli spin matrices $\sigma_x$ and $\sigma_z$.

A general Pauli (also called Heisenberg-Weyl) operator is of the form

\begin{equation}
T_{(m,j)} = \left\{
    \begin{array}{ll}
        i^{jm}Z^m X^j & \mbox{if } d=2 \\
        \omega^{-jm/2}Z^m X^j & \mbox{if } d \ne 2.
    \end{array}
\right.
\label{eqn2}
\end{equation}
where $(j,m) \in \mathbb{Z}_d \times \mathbb{Z}_d$.
For $N$ particules, one takes the Kronecker product of qudit elements $N$ times.

Stabilizer states are defined as eigenstates of the Pauli group.

\subsection*{The discrete Wigner function}
\label{DWF}
Associated with each $d$-dimensional Hilbert space ($d$ a prime) is a discrete phase space, a $d \times d$ array of points on $\mathbb{Z}_d \times \mathbb{Z}_d$. 
A set of phase point operators on the discrete phase space is defined as \cite{Wootters1987, Ferrie2011} (see also \cite{Revzen2016})
\begin{equation}
A_{\alpha}=\frac{1}{d}\sum_{j,m=0}^{d-1}\omega^{pj-qm+jm/2}X^jZ^m,~~\alpha=(q,p) \in \mathbb{Z}_d \times \mathbb{Z}_d .
\label{eqn3}
\end{equation}
Wootters relations (10) and (11) in \cite{Wootters1987} follow. 

Phase point operators have been built to satisfy properties analogous to those of the continuous phase space in the context of the continuous Wigner function 
$W(q,p)=\int \rho(q+x/2,q-x/2)\exp(ipx) dx$ in which $p$ and $q$ are position and momentum, and $\rho(x,x')=\psi^*(x)\psi(x)$ is a density matrix for a particle of coordinate $x$ in a pure state of wave function $\psi(x)$ \cite[p. 477]{Feynman1982}. It is required that the operators $A_{\alpha}$ satisfy

(i) For each point $\alpha$, $A_{\alpha}$ is Hermitian,

(ii) For any two points $\alpha$ and $\beta$, $\mbox{tr}(A_{\alpha} A_{\beta})=d\delta_{\alpha\beta}$,

(iii)Taking any complete set of $d$ parallel lines (called a striation), construct the average $P_{\lambda}=\frac{1}{d}\sum_{\alpha \in \lambda}A_{\alpha}$ on each line $\lambda$. The $d$ operators $P_{\lambda}$ form a set  of mutually orthogonal projectors the sum of which is the identity operator.

The $d^2$ phase point operators $A_{\alpha}$ are linearly independent and form a basis for the space of Hermitian operators acting on a $d$-dimensional Hilbert space. As a result, any density operator can be developed as 
\begin{equation}
\rho=\sum_{q,p}W_{\rho}(q,p)A(q,p),
\label{eqn4}
\end{equation}
in which the real coefficients are explicitely given by the Wootters discrete Wigner function
\begin{equation}
W_{\rho}(q,p)=\frac{1}{d}\mbox{tr}[\rho A(q,p)].
\label{eqn5}
\end{equation}
Unlike the continuous Wigner function, the discrete Wigner function is a quasi probability distribution that may take negative values. It is shown in \cite{Gross2007} that, on a Hilbert space of odd dimension, the only pure states to possess a non-negative discrete Wigner function are stabilizer states.

On the contrary, a non-stabilizer pure state will be called a magic state. This definition follows from \cite[Corollary 1]{Reichardt2005} which establishes that any single-qubit pure state not one of the six Pauli eigenstates, together with Clifford group operations and Pauli eigenstate preparation and measurement, allows universal quantum computation. For arbitrary prime dimensions, magic state distillation is investigated in \cite{Meier2012}-\cite{Veitch2012}.
 For multiple qubits see \cite[Corollary 2]{Reichardt2005} any pure state which is not a stabilizer state allows universal quantum computation and \cite{Raussendorf2016}.

\subsection*{State-dependent quantum contextuality}
Quantum contextuality forbids theories revealing pre-existing values of observables under test if the specific experimental set-up for measuring such observables is not taken into account. One way to characterize quantum contextuality is to use a no-go approach a la Kochen-Specker involving sets of quantum observables (as the Mermin square) or special subsets of their eigenstates shared by their mutually commuting operators \cite{Planat2012}. The smallest proof of state-independent contextuality is not a Kochen-Specker set, it needs $13$ rays in the three-dimensional Hilbert space \cite{Cabello2016}. State-independent contextuality can be obtained within the stabilizer formalism for multiple qubits but is not manifested with qudits (when $d \ne 2$).

The starting point of a state-dependent proof of quantum contextuality consists of a set of $k$ binary tests that are represented by $k$ rank one projectors
 $\{ \Pi_i=\left|v_i \right \rangle\left \langle v_i \right |, i=1..k\}$. The tests are compatible if and only if the projectors are mutually orthogonal. To the tests is associated an (exclusivity) graph $\Gamma$ wherein each vertex corresponds to the projectors $\Pi_i$ and an edge corresponds to compatible projectors. A witness operator $\Sigma_{\Gamma}$ is defined as follows
\begin{equation}
\Sigma_{\Gamma}=\sum_{i=1}^k \Pi_i.
\end{equation}
It is required that a value of $1$ is assigned to at most one projector in each joint measurement of (compatible) observables located on a selected edge.
For a non-contextual hidden variable theory, one expects that the results of tests is such that $\left \langle \Sigma_{\Gamma}\right\rangle_{\mbox {max}} \le \alpha(\Gamma)$, where $\alpha(\Gamma)$ is the independence number of the graph $\Gamma$, the cardinality of the largest set of vertices such that no two elements are connected by an edge. But a quantum contextual theory may bypass this bound and be such that $\alpha(\Gamma) \le \left \langle \Sigma_{\Gamma} \right \rangle_{\mbox {max}}  \le\theta(\Gamma)$, in which the upper bound is the Lovasz number of the exclusivity graph \cite{Howard2014,Cabello2010}. It is calculated from
%
$\theta(G)=\mbox{max}_{\left|\psi\right \rangle} \sum_{i=1}^k \left| \left \langle \psi |v_i\right \rangle \right|^2$,
%
where the maximum is taken over all unit vectors $\left|\psi\right \rangle$. 

The simplest state-dependent proof of quantum contextuality corresponds to the cyclic graph $C_5$ (also called a pentagon) in which $\theta(C_5)=\sqrt{5}>\alpha(C_5)=2$. It was originally obtained for a spin-$1$ quantum system, or qutrit \cite{Klyachko2008}. 
Each exclusivity graph $\Gamma$ where $\alpha(\Gamma)<\theta(\Gamma)$ allows a state-dependent proof of quantum contextuality. If $\left \langle \Sigma_{\Gamma} \right \rangle_{\mbox {max}}>\alpha(\Gamma)$ for every state, then the proof of contextuality is state-independent. To summarize, graphs in which $\alpha(\Gamma)<\theta(\Gamma)$ may be considered a proof of quantum contextuality if for appropriately chosen projectors $\Pi_i$ and a state $\rho$ the following non-contextuality inequality is violated \cite{Howard2012}.
\begin{equation}
\mbox{Tr}(\Sigma_{\Gamma} \rho)\le \alpha(\Gamma). 
\label{eqn7}
\end{equation}

Is contextuality needed for universality and quantum computational speed-up? For qudits (in odd dimension), only states lying outside the stabilizer polytope \cite{Cormick2006} manifest the negativity of the Wigner function and simultaneously violate the inequality (\ref{eqn7}) through appropriate two-qudit projectors, hence they display state-dependent contextuality \cite[Theorem 1]{Howard2014}. Thus the answer is yes.

For even dimensions, non-stabilizer pure states of single or multiple qubits are magic. But neither the condition of negativity nor the criterion of contextuality is sufficient to promote such states to computational universality. This is because there exist (well characterized) Kochen-Specker sets of multi-qubits \cite{Planat2015}. However, it is shown in a recent paper \cite{Raussendorf2016} that contextuality is needed whenever two conditions are satisfied, (i) the contextuality is state-dependent and (ii) one retains a filtered set of quantum states able to ensure a full state tomography.

\section{Magic states and state-dependent quantum contextuality from groups of permutation gates}
We already know that permutation symmetry exists in the discrete Wigner function \cite{Zhu2016}. Our goal in this section is to shift our attention from the set of Clifford gates to a subset whose elements are permutation gates, leading either to stabilizer states or not. Classes of magic states and the related (state-dependent) contextuality will be investigated having in mind the tools described in Sec. \ref{Sec2} but by restricting to the existence of pentagons between the appropriate states (as in \cite{Klyachko2008}). 

For dimensions larger than $4$, the eigenstates of permutation gates under consideration are living in a field that may be different from the cyclotomic field $\mathbb{Q}[\exp\frac{2i\pi}{d}]$. In this paper, for $d>4$, we restrict the classification of magic states to those having entries $0$ and $\pm 1$ corresponding to eigenvalues $\pm 1$.

\subsection{Qubit magic states}
Fault tolerant quantum computing protocols based on stabilizer states have to be complemented by magic states to reach quantum universality. Two distillation protocols based on single qubit magic states $\left| H\right\rangle$ and  $\left| T\right\rangle$ are first described in  \cite{Bravyi2004} where
\begin{eqnarray}
&\left| H\right\rangle=\cos(\frac{\pi}{8})\left| 0\right\rangle + \sin(\frac{\pi}{8})\left| 1\right\rangle, \\
&\left| T\right\rangle=\cos(\beta)\left| 0\right\rangle + \exp{(\frac{i\pi}{4})}\sin(\beta)\left| 1\right\rangle,~~ \cos(2\beta)=\frac{1}{\sqrt{3}}.\\
\nonumber
\end{eqnarray}
Magic state $\left| H\right\rangle$ is the $+1$-eigenstate of the the Hadamard matrix $H=\frac{\sigma_x+\sigma_z}{\sqrt{2}}$ that belongs to single qubit the Clifford group. 
Magic state $\left| T\right\rangle$ is the $\omega_3=\exp(\frac{2i\pi}{3})$-eigenstate of the $SH$ matrix, where the phase gate $S=\bigl( \begin{smallmatrix} 
  1 & 0\\
  0 & i 
\end{smallmatrix}\bigr)$ also belongs to the Clifford group. Being not stabilizer states, they cannot be prepared by actions from the Clifford group but they can be distilled with these actions from a (dirty) mixed state to a (neat) pure state thanks to appropriate quantum codes \cite{Bravyi2004, Reichardt2005, Meier2012}.

\begin{table}[t]
\begin{center}
\begin{tabular}{|l|c|r|}
\hline 
state $\psi$ & eigenstate of & $W_{\rho} $\\
\hline
$\left| 0\right\rangle $ & $\sigma_z$ &
 $\frac{1}{2}\bigl( \begin{smallmatrix} 
  1 & 1\\
  0 & 0 
\end{smallmatrix}\bigr)$ \\
\hline
$\frac{1}{\sqrt{2}}(\left| 0\right\rangle+\left| 1\right\rangle)$ &$\sigma_x$ &
$\frac{1}{2}\bigl( \begin{smallmatrix} 
  1 & 0\\
  1 & 0 
\end{smallmatrix}\bigr)$ \\
\hline
$\frac{1}{\sqrt{2}}(\left| 0\right\rangle-i\left| 1\right\rangle)$ &$\sigma_y$ &
$\frac{1}{2}\bigl( \begin{smallmatrix} 
  0 & 1\\
  1 & 0 
\end{smallmatrix}\bigr)$ \\
\hline
$\left| H\right\rangle$ & $H=\frac{\sigma_x+\sigma_z}{\sqrt{2}}$&
$\frac{1}{4}\bigl( \begin{smallmatrix} 
  1+\sqrt{2} & 1\\
  1 & 1-\sqrt{2} 
\end{smallmatrix}\bigr)$ \\
\hline
$\left| T\right\rangle$ & $SH$&
$\frac{1}{4}\bigl( \begin{smallmatrix} 
  1+\sqrt{3}/3 &  1+\sqrt{3}/3\\
   1+\sqrt{3}/3 & \sqrt{3}-1 
\end{smallmatrix}\bigr)$ \\
\hline
\end{tabular}
\caption{Wootters discrete Wigner function for a few pure states $\rho=\left|\psi \right\rangle \left\langle \psi \right|$.}
\end{center}
\end{table}
Matrix elements of Wootters discrete Wigner function $W_{\rho}(q,p)$ for a few single qubit pure states are shown in Table 1. The first three rows correspond to $+1$-eigenstates of Pauli spin matrices and the last two rows are for magic states. Magic state $\left| H\right\rangle$ possesses a negative entry unlike the magic state $\left| T\right\rangle$. In the latter case, negativity of the Wigner function does not coincide to universality.

\subsection{Magic states from qutrit permutation gates}
The smallest dimension for the occurrence of magic states associated to permutations is three. There are, up to isomorphism, two (non-trivial, i.e. with two distinct generators) two-generator permutation groups on three letters. 

The permutation group isomorphic to $\mathbb{Z}_3$ contains the permutation matrices $I$, $X$, and $X^2$ of the Pauli group, where $X$ is the shift matrix in (\ref{eqn1}). The eigenstates are the mutually orthogonal stabilizer states $(1,1,1)$, $(1,\omega,\omega^2)$ and $(1,\omega^2,\omega)$, with $\omega$ the third root of unity.

\begin{figure}[ht]
\includegraphics[width=6cm]{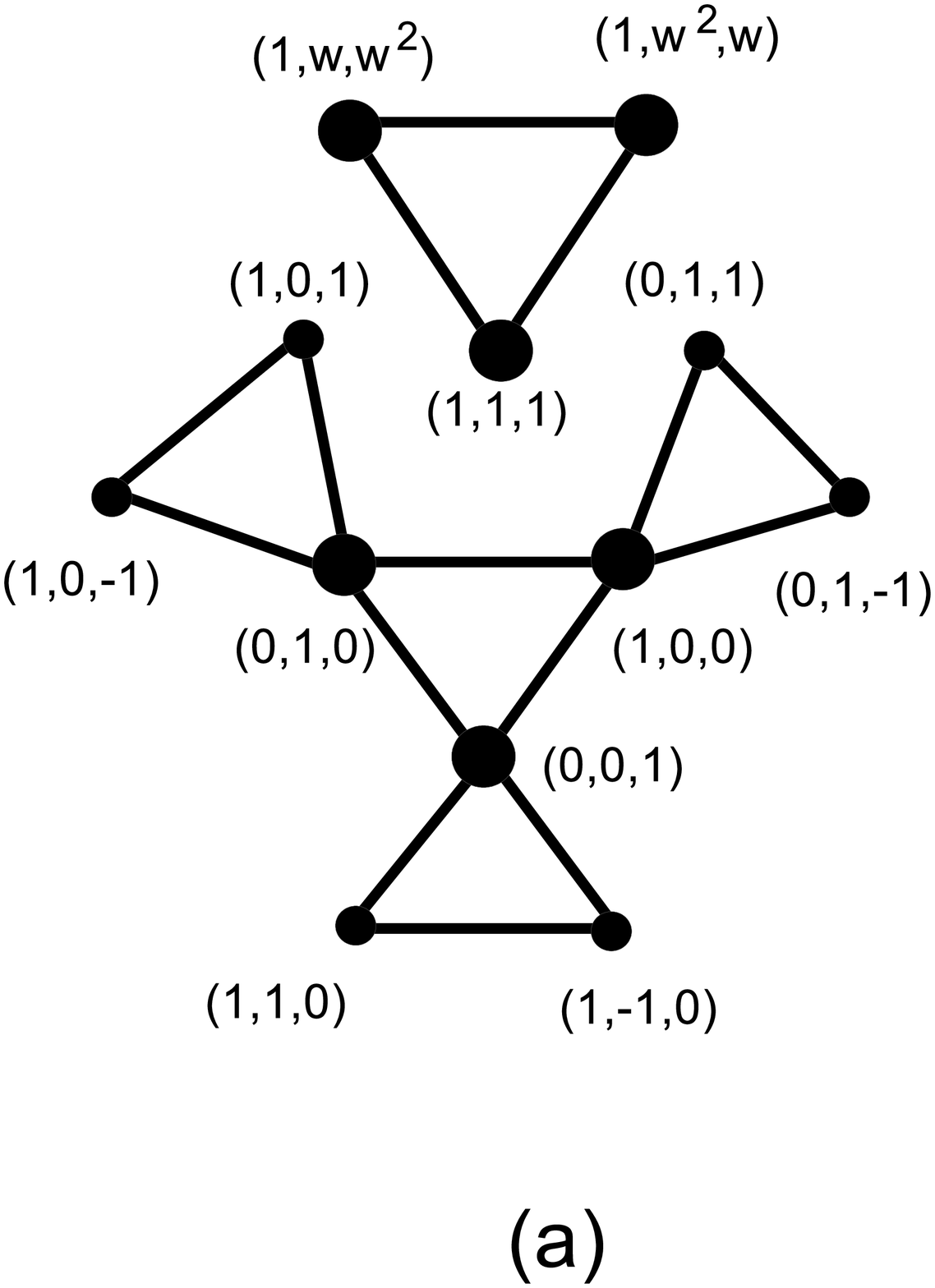}
\includegraphics[width=6cm]{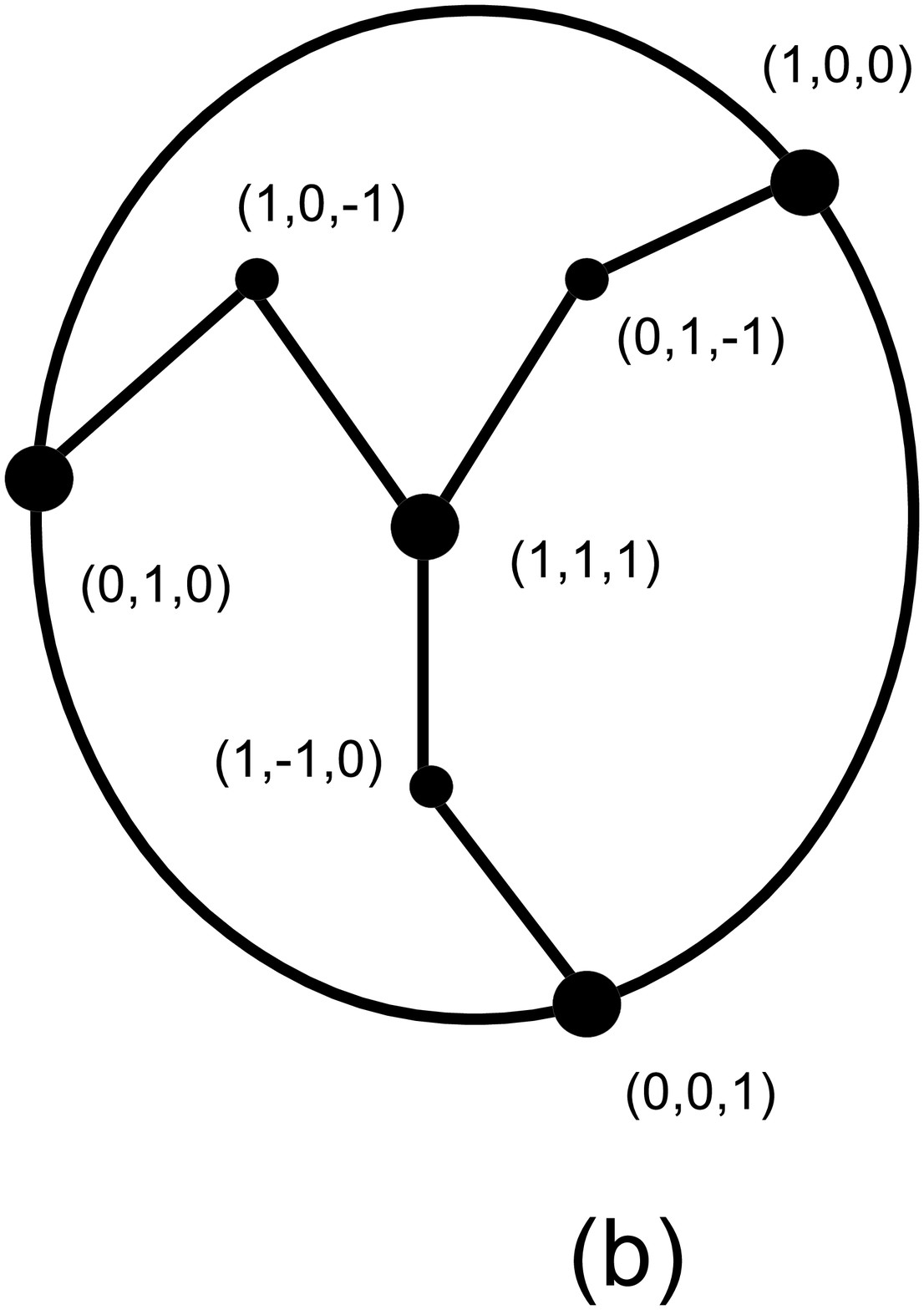}
\caption{(a) Triples of mutually orthogonal rays arising from qutrit permutation gates, (b) three pentagons originating from the rays. Big black bullets are for stabilizer states, small bullets are for magic states.
 }
\end{figure} 

The permutation group isomorphic to the symmetric group $S_3$ exists in three copies. One of them is generated as $S_3=\left\langle (1,2,3), (2,3)\right\rangle$ and contains the elements found in $\mathbb{Z}_3$ and three extra ones $\bigl( \begin{smallmatrix} 
  1 & 0&0\\
  0&0 & 1 \\
	0&1&0
\end{smallmatrix} \bigr)\equiv (2,3)$, $\bigl( \begin{smallmatrix} 
  0 & 0&1\\
  0&1 & 0 \\
	1&0&0
\end{smallmatrix} \bigr)\equiv (1,3)$ and $\bigl( \begin{smallmatrix} 
  0 & 1&0\\
  1&0 & 0 \\
	0&0&1
\end{smallmatrix} \bigr)\equiv (1,2)$, that do not lie in Pauli group but are parts of the Clifford group. The $5$ mutually orthogonal triples between the $12$ eigenstates are pictured in Fig. 1a. Apart from the $6$ stabilizer states, there are $6$ magic states of type $(0,1,1)$ and $(0,1,-1)$. In \cite{Veitch2014}, they are identified as the Norrell states and strange states, respectively, with Wigner function.
\begin{equation}
W_{(0,1,1)}=\frac{1}{6}\left( \begin{smallmatrix} 
  2 & -1&-1\\
  1&1 & 1 \\
	1&1&1
\end{smallmatrix} \right),~~W_{(0,1,-1)}=\frac{1}{6}\left(\begin{smallmatrix} 
  -2 & 1&1\\
  1&1 & 1 \\
	1&1&1
\end{smallmatrix} \right).
\end{equation}
As expected, all magic states contains some negative entries in their Wigner matrix.

In Fig 1b, it is shown that three pentagons are part of the orthogonality relations. Taking the rank $1$-projectors associated to the vertices/states, the exclusivity graph attached to each pentagon is also a pentagon and allows a state-dependent proof of contextuality. One observes that only magic states of the strange type are involved.

\subsection{Magic states from two-qubit permutation gates}
From now we restrict to permutation groups whose two generators are magic gates. This only happens for two groups both isomorphic to the alternating group $A_4$. One copy is as follows
\begin{equation}
A_4\cong \left\langle   \left(\begin{smallmatrix} 
  1 & 0&0&0\\
  0&0 &0& 1 \\
	0&1&0&0\\
	0&0&1&0
\end{smallmatrix} \right),
 \left( \begin{smallmatrix} 
  0 & 1&0&0\\
  0&0 &1& 0 \\
	1&0&0&0\\
	0&0&0&1
\end{smallmatrix} \right) \right\rangle.
\label{eqn11}
\end{equation}

\begin{figure}[ht]
\includegraphics[width=6cm]{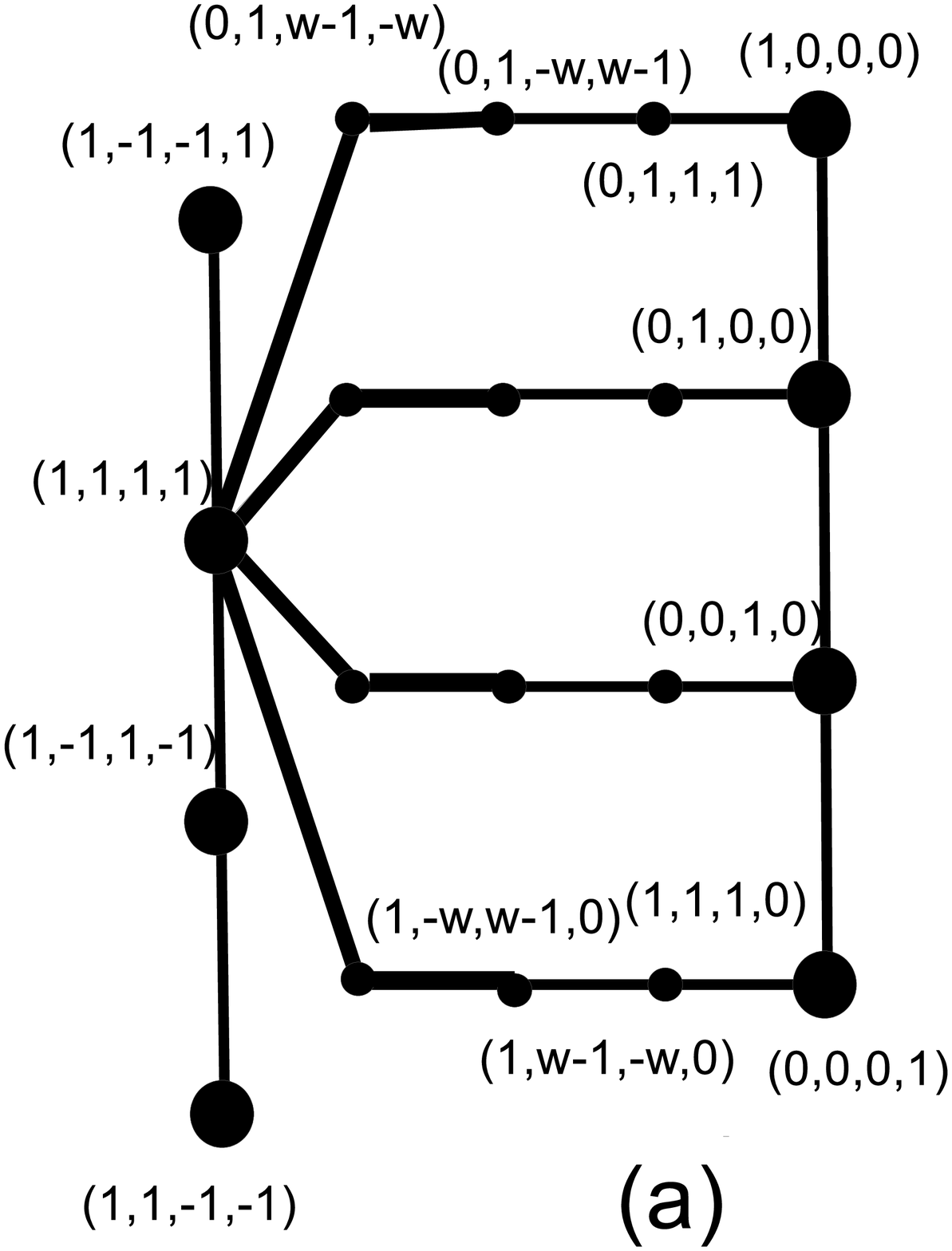}
\includegraphics[width=6cm]{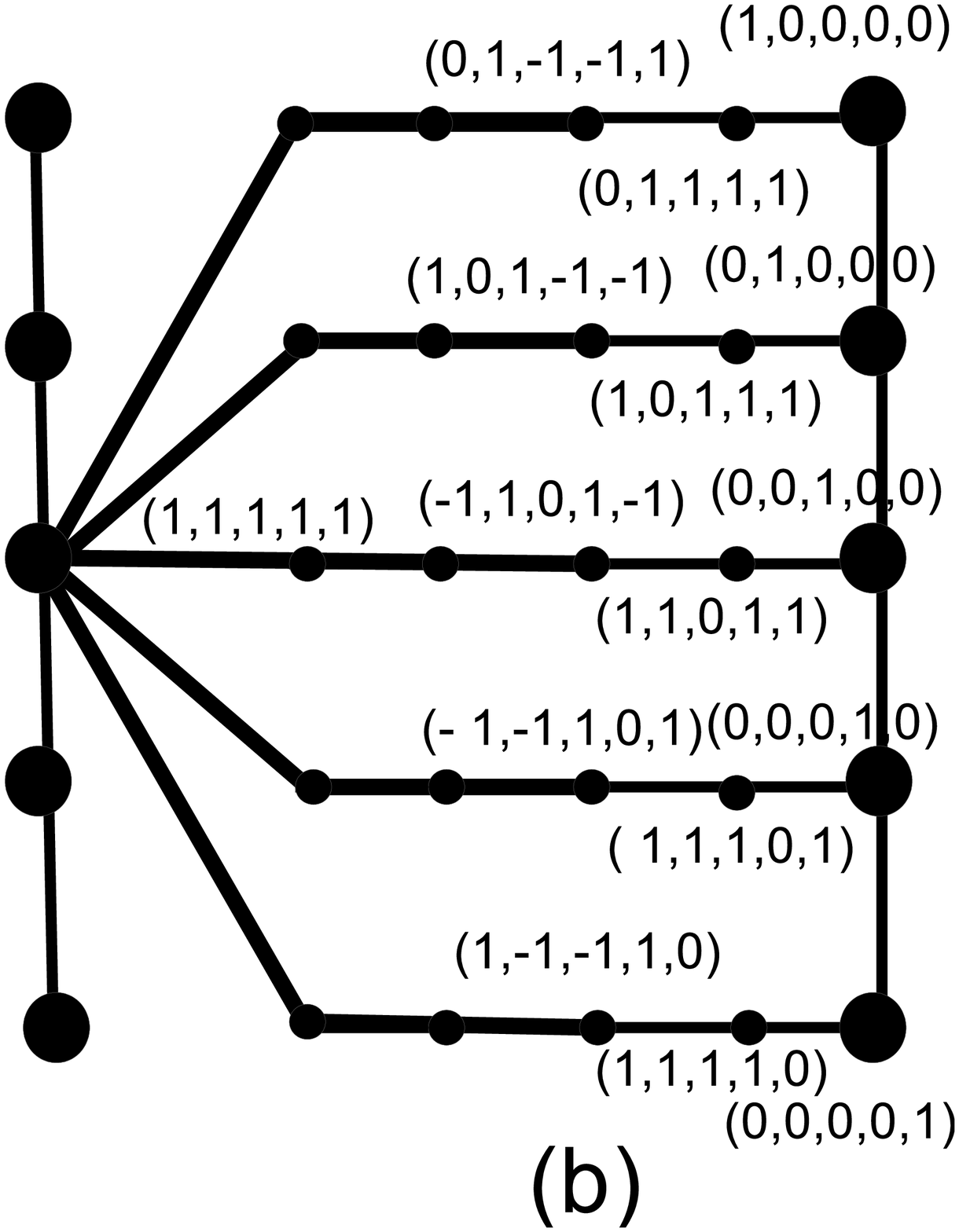}
\caption{(a) Maximum cliques of orthogonal rays from two-qubit permutation gates: $6$ $4$-tuples (thin lines) and $4$ triples (thick lines), $\omega=\exp(\frac{i\pi}{3})$, Missing coordinates are straightforward to recover.
(b) Maximum cliques of orthogonal rays from $5$-dit permutation gates: $7$ $5$-tuples (thin lines) and $5$ $4$-tuples (thick lines). Big black bullets are for stabilizer states, small black bullets are for magic states.
 }
\end{figure} 

Looking for joined eigenstates shared by at least two commuting gates, a set of $20$ states is derived whose orthogonality graph is pictured in Fig. 2a.

Wigner functions for magic states are of two types
\begin{equation}
W_{(0,1,1,1)}=\frac{1}{12}\left(\begin{smallmatrix} 
  1 & -1&2&2\\
  -1&1 & 0&0 \\
	2&0&4&0\\
	2&0&0&0
\end{smallmatrix} \right),~~W_{(0,1,-\omega,\omega-1)}=\frac{1}{24}\left(\begin{smallmatrix} 
  -1 & 1&1& 1+2\sqrt{3}\\
  1&-1 & 3& 3-2\sqrt{3}\\
	1&3&-1&3-2\sqrt{3}\\
	1-2\sqrt{3}&3+2\sqrt{3}&3+2\sqrt{3}&3
\end{smallmatrix} \right),
\end{equation}
with $W_{(0,1,\omega-1,-\omega)}=W_{(0,1,-\omega,\omega-1)}^t$. Both types of Wigner matrices  contain negative entries as for the qutrit case (while in this particular case, the stabilizer states selected from permutation gates do not).

It can be shown that the orthogonality graph contains $24$ pentagons and thus $24$ state-dependent proofs of contextuality. Surprisingly, the vertices of such pentagons either are stabilizer states or magic states of the second type, analogous to qutrit strange states. 

\subsection{Magic states from $5$-dit permutation gates}
We again restrict restrict to permutation groups whose two generators are magic gates. This happens for permutation groups isomorphic to the semidirect product $\mathbb{Z}_5 \rtimes \mathbb{Z}_4$ or to the symmetric group $S_5$, respectively.

As for the first case, taking all triples of mutually commuting/compatible permutation gates, one gets a set of $30$ eigenstates. They are organized into maximum cliques shown in Fig. 2b with $20$ of them being magic. Missing coordinates are in the cyclotomic field $\mathbb{Q}[\exp(\frac{2i\pi}{4})]$.


%
\begin{equation}
W_{(0,1,1,1,1)}=\frac{1}{20}\left(\begin{smallmatrix} 
  4 & -1&-1&-1&-1\\
  3&+ & -'&-'&+ \\
	3&-'&+&+&-'\\
	3&-'&+&+&-'\\
	3&+&-'&-'&+\\
	\end{smallmatrix} \right),
	~~
	W_{(0,1,-1,-1,1)}=\frac{1}{20}\left(\begin{smallmatrix} 
  4 & -1&-1&-1&-1\\
  -1&-'' & +''&+''&-'' \\
	-1&+''&-''&-''&+''\\
	-1&+''&-''&-''&+''\\
	-1&-''&+''&+''&-''\\
	\end{smallmatrix} \right),
\end{equation}
where the symbol \lq$\pm$' means $\pm(1+ \sqrt{5})/2$, the symbol \lq$\pm'$' means $\pm(\sqrt{5}-1)/2$
 and \lq$\pm ''$' means $(3 \pm\sqrt{5})/2$.

As for the group $S_5$, one gets $50$ eigenstates (shared by the $10$ maximum cliques of $5$ mutually compatible gates).
We do not show the organization of eigenstates but only mention that one arrives at other types of magic states of the form


%
\begin{equation}
W_{(0,0,0,1,\pm 1)}=\frac{1}{10}\left(\begin{smallmatrix} 
  0 & 0&0&0&0\\
  -2&\pm' & \mp&\mp&\pm' \\
	0&0&0&0&0\\
	1&1&1&1&1\\
	1&1&1&1&1\\
	\end{smallmatrix} \right),
	~~
	W_{(0,0,1,1,1)}=\frac{1}{15}\left(\begin{smallmatrix} 
  2 & +'&-&-&+'\\
  2&+' & -&-&+' \\
	1&1&1&1&1\\
	3&-'&+&+&-'\\
	1&1&1&1&1\\
	\end{smallmatrix} \right).
\end{equation}

As before, state-dependent contextuality may be revealed from the pentagons that are built from the states.

\subsection{Magic states from qubit-qutrit permutation gates}
The smallest permutation group generated by two magic permutation gates in dimension $6$ is the alternating group $A_5$. There exist maximum cliques 
of mutually compatible permutation gates whose size is $2$, $3$ and $4$ giving rise to shared eigenstates (stabilizer and magic) defined over the cyclotomic field $\mathbb{Q}[\exp(\frac{2i\pi}{6})]$.
As expected for even dimensions, the negativity in the entries of the Wigner function is not the sign of magicity of the state, e.g. for the stabilizer state $(1,-1,1,-1,1,-1)$ one gets
\begin{equation}
W_{(1,-1,1,-1,1,-1)}=\frac{1}{18}\left(\begin{smallmatrix} 
 0 & 0&0&0&0&0\\
  0&0 &0&-1&2&2 \\
	3&0&0&-1&2&2\\
	0&0&0&0&0&0\\
	0&0&0&-1&2&2\\
	3&0&0&-1&2&2
	\end{smallmatrix} \right).
	\end{equation}
 A representative magic state
 is $(0,1,1,1,1,1)$ with Wigner function
\begin{equation}
W_{(0,1,1,1, 1, 1)}=\frac{1}{30}\left(\begin{smallmatrix} 
 4 & -2&-2&3&+&-\\
  3&- &+&0&0&0 \\
	-1&-''&+'&-1&+'&-''\\
	5&-1&-1&5&-''&+'\\
	5&+'&-''&1&1&1\\
	1&+''&-'&1&-'&+''
	\end{smallmatrix} \right),
	\label{eqn16}
	\end{equation}
	where in (\ref{eqn16}) the notation\lq$\pm$' means $(3 \pm \sqrt{3})/2$, \lq$\pm'$' means $\pm (\sqrt{3}+1)/2$ and 
\lq$\pm''$' means $\pm (\sqrt{3}-1)/2$.

Another permutation group generated by two magic permutation gates is the alternating group $A_6$ giving rise to magic states such as $(0,0,1,1,\pm 1,\pm 1)$.
One gets
\begin{equation}
W_{(0,0,1,1,1,1)}=\frac{1}{24}\left(\begin{smallmatrix} 
 1 & -&+'&1&+'&-\\
 2&2 &2&-1&+&-' \\
	-1&-'&+&0&0&0\\
	4&+'&-&4&-&+'\\
	4&1&1&2&-'&+\\
	2&+&-'&2&-1&-1
	\end{smallmatrix} \right),
	\label{eqn17}
	\end{equation}
	where in (\ref{eqn17}) the notation \lq$\pm$' means $ \pm (1+\sqrt{3})/2$ and \lq$\pm'$' means $\pm (\sqrt{3}-1)/2$.

\subsection{Magic states from $7$-dit permutation gates}
The smallest permutation group generated by two magic permutation gates in dimension $7$ is isomorphic to $\mathbb{Z}_7 \rtimes \mathbb{Z}_6$.
 A representative magic state is $(0,1,1,1,1,1,1)$ with Wigner function
\begin{equation}
W_{(0,1,1,1,1,1,1)}=\frac{1}{42}\left(\begin{smallmatrix} 
 6 & -1&-1&-1&-1&-1&-1\\
 5 & b&c&-a&-a&c&b\\
5 &c&-a&b&b&-a&c\\
5 & -a&b&c&c&b&-a\\
5 & -a&b&c&c&b&-a\\
5 &c&-a&b&b&-a&c\\
5 & b&c&-a&-a&c&b
	\end{smallmatrix} \right),
	\end{equation}
where $a=2\cos(2\pi/7)$, $b=-2\cos(4\pi/7)$ and $c=-2\cos(6\pi/7)$ are positive so that a negative sign in front of the matrix entries is a negative entry of the Wigner function.

Next, another permutation group generated by two magic gates is isomorphic to $PSL(2,7)$, of order $168$.
 One finds three types of magic states whose entries are $0$ or $\pm1$ with Wigner functions as follows
\begin{eqnarray}
&W_{(0,0,0,0,1,1,1)}=\frac{1}{21}\left(\begin{smallmatrix} 
 0 & 0&0&0&0&0&0\\
 2 & a&-b&-c&-c&-b&a\\
2 & a&-b&-c&-c&-b&a\\
0 & 0&0&0&0&0&0\\
1 & 1&1&1&1&1&1\\
3 &1-b&1-c&1+a&1+a&1-c&1-b\\
1 & 1&1&1&1&1&1
	\end{smallmatrix} \right), \nonumber \\
	%
%
&W_{(0,0,0,0,0,1,\pm 1)}=\frac{1}{14}\left(\begin{smallmatrix} 
 0 & 0&0&0&0&0&0\\
0 & 0&0&0&0&0&0\\
\pm 2 & \pm a&\mp b&\mp c&\mp c&\mp b&\pm a\\
0 & 0&0&0&0&0&0\\
0 & 0&0&0&0&0&0\\
1 & 1&1&1&1&1&1\\
1 & 1&1&1&1&1&1
	\end{smallmatrix} \right), \nonumber \\
	%
%
&W_{(0,0,0,1,1,\pm 1,\pm 1)}=\frac{1}{28}\left(\begin{smallmatrix} 
 2 & -a&-b&-c&-c&-b&a\\
4 & \pm(b-1)&\pm(c-1)&\mp(1+a)&\mp(1+a)&\pm(c-1)&\pm(b-1)\\
2 & -a&-b&-c&-c&-b&a\\
1 & 1&1&1&1&1&1\\
1&b\mp 1&1 \mp c& 1 \pm a& 1 \pm a& 1 \mp c& b \mp 1\\
1&b\mp 1&1 \mp c& 1 \pm a& 1 \pm a& 1 \mp c& b \mp 1\\
1 & 1&1&1&1&1&1
	\end{smallmatrix} \right),
	\end{eqnarray} \nonumber \\
where $1-b>0$, $1-c<0$ and $1-a<0$.

\subsection{Higher dimensions}

Table 2 summarizes the magic states found from permutation gates of small dimensions $d \le 9$. For dimensions $d\ge 5$, only magic states with entries $0$ or $\pm 1$ are considered. Column 3 provides the   sum of negative entries in the Wigner matrix (\ref{eqn5}). It is shown in \cite{Veitch2014} that the absolute value $\mbox{sn}(\rho)$ of the sum of negative entries in the discrete Wigner matrix $W_{\rho}$ is a computable magic monotone. It is a quantum computing resource that does not increase under stabilizer operations. Similarly the so-called mana $\mathcal{M}(\rho)=\log[(2\mbox{sn}(\rho])$ is an (easily computable) additive magic monotone. According to \cite[Theorem. 14]{Veitch2014}, a stabilizer protocol succeeds probabilistically to produce $m$ copies of the target state $\sigma$ from at least $m\frac{\mathcal{M}(\sigma)}{\mathcal{M}(\rho)}$ copies of the state $\rho$ on average.

\footnotesize
\begin{table}[t]
\begin{center}
\begin{tabular}{|l|c|r|l|}
\hline 
dim& magic state $\rho$& sum of negative entries $W_{\rho}$ & Remark\\
\hline
2 & $\left| H \right \rangle$ & $(1-\sqrt{2})/4 \sim -0.1035$ & \cite{Bravyi2004}\\
  & $\left| T \right \rangle$ & positive & \cite{Bravyi2004} \\
\hline
3 & $(0,1,1)$ & -1/3 & Norrell \cite{Veitch2012} \\ 
 & $(0,1,-1)$ & -1/3 & strange \cite{Veitch2012} \\
\hline
4 & $(0,1,1,1)$ & $-1/6$& $A_4$ \\
 & $(0,1,-\omega,\omega-1)$ &$(2-3 \sqrt{3})/12 \sim -0.266$ & \\
\hline
5 & $(0,1,1,1,1)  $ & $- \sqrt{5}/5 \sim -0.447$      & $\mathbb{Z}_5 \rtimes \mathbb{Z}_4$\\
  & $(0,1,-1,-1,1)$ & $ -2/5$ & \\
    & $(0,0,0,1,\pm 1)$  & $-(\sqrt{5}+1)/10\sim -0.324$ & $S_5$ \\
  & $(0,0,1,1,1)$   & $-(1+3 \sqrt{5})/15\sim -0.514$ &   \\
\hline
6 & $(0,1,1,1,1,1)$ & $-(3 \sqrt{3}+7)/30 \sim -0.406$ & $A_5$ \\
 & $(0,0,1,1,1,1)$ & $-(\sqrt{3}+1)/6 \sim -0.455$ & $A_6$ \\
& $(0,0,1,-1,-1,1)$ & $-(\sqrt{3}+4)/12 \sim -0.478$ &  \\
\hline
7 & $(0,1,1,1,1,1,1)$ & $-[1+4\cos(2\pi/7)]/7 \sim -0.499$ & $\mathbb{Z}_7 \rtimes \mathbb{Z}_6$ \\
  & $(0,0,0,0,1,1,1)$ & $[2+8 \cos(4\pi/7)+12\cos(6\pi/7)]/21\sim -0.504$ & $PSL(2,7)$\\
	& $(0,0,0,0,0,1,\pm 1)$ & $2[\cos(4\pi/7)+\cos(6\pi/7)]/7\sim -0.321$ &\\
	& $(0,0,0,1,1,1,1)$ & $[1+4 \cos(4\pi/7)+10\cos(6\pi/7)]/14\sim -0.636$ &\\
	& $(0,0,0,1,1,-1,-1)$ & $[2+8 \cos(4\pi/7)+10\cos(6\pi/7)]/14\sim -0.628$ &\\
	\hline
8 & $(0,1,1,1,1,1,1,1)$ & $-37/112 \sim -0.330$ & $\mathbb{Z}_2^3 \rtimes \mathbb{Z}_7$ \\
\hline	
9 & $(0,0,0,0,0,1,1,1,1)$ & $-1/2 $ & $\mathbb{Z}_3^2 \rtimes \mathbb{Z}_4$ \\
 & $(0,1,1,1,1,1,1,1,1)$ & $-5/9 \sim -0.555$ & $\mathbb{Z}_3^2 \rtimes \mathbb{Z}_8$ \\
& $(0,0,0,0,1,1,1,1,1)$ & $-2/3$ & $G_{144}$ \\
& $(0,0,0,0,0,0,0,1,1)$ & $-1/3$ & \\
& $(0,0,0,1,-1,1,-1,1,-1)$ & $-19/27\sim -0.704$ &  \\
\hline
\end{tabular}
\caption{The magic states $\rho$ (column 2) and the sum of negative entries in the Wigner matrix $W_{\rho}$ (column 3). For $d>3$, column 4 provides the permutation group under consideration. The group $G_{144}$ is isomorphic to $\mathbb{Z}_3^2 \rtimes(\mathbb{Z}_2 \rtimes D_4)$, where $D_4$ is the eight element dihedral group.}
\end{center}
\end{table}

\normalsize
\section{Conclusion}

We described the leading role played by permutations in shaping a type of universal quantum computation based on magic states -states living outside the stabilizer polytope defined by the (generalized) Pauli group. In Sec. 3, we derived the main magic states $\rho$ defined from eigenstates of gates in magic permutation groups (a subset of Clifford group) of small dimensions $d\le 9$. We explicitly computed the sum negativity of the discrete Wigner matrix $W_{\rho}$  for estimating the value of a magic state as a resource for universal quantum computation \cite{Veitch2012}. We observed that state dependent quantum contextuality (through building block pentagons) occurs in all dimensions $d \ge 3$ from appropriate sets of stabilizer and magic states. It is desirable to extend the calculations to higher dimensions $d>9$ to check a possible asymptotic trend between the dimension and the amount of magicity and contextuality, to clarify the relation between magic permutation groups and unitary $2$-designs (singled out in \cite{Zhu2016})   and finally to relate the derived magic states to distillation procedures and error correcting codes. A later paper will be devoted to the POVMs obtained from the Pauli group action on the magic states \cite{Planat2017}.


\end{document}